\documentclass[12pt,3p,sort&compress]{elsarticle}

\pdfoutput=1

\usepackage{url}
\usepackage{graphicx}
\usepackage{amssymb}
\usepackage[usenames]{color}

\begin{document}

\title{Anisotropic Hydrodynamics for Rapidly Expanding Systems}

\begin{frontmatter}

\author[kielce,krakow]{Wojciech Florkowski} 
\author[krakow]{Radoslaw Ryblewski} 
\author[kent,frankfurt]{Michael Strickland} 
\address[kielce]{Institute of Physics, Jan Kochanowski University, PL-25406~Kielce, Poland}
\address[krakow]{The H. Niewodnicza\'nski Institute of Nuclear Physics, Polish Academy of Sciences, PL-31342 Krak\'ow, Poland}
\address[kent]{Department of Physics, Kent State University, Kent, OH 44242 United States}
\address[frankfurt]{Frankfurt Institute for Advanced Studies, Ruth-Moufang-Strasse 1, D-60438, Frankfurt am Main, Germany}

\begin{abstract}
We exactly solve the relaxation-time approximation Boltzmann equation for a system which is transversely homogeneous and undergoing boost-invariant longitudinal expansion.  
We compare the resulting exact numerical solution with approximate solutions available in the anisotropic hydrodynamics and second order viscous hydrodynamics frameworks.  
In all cases studied, we find that the anisotropic hydrodynamics framework is a better approximation to the exact solution than traditional viscous hydrodynamical approaches.
\end{abstract}


\end{frontmatter}


The application of relativistic viscous hydrodynamics is important in a wide variety of situations including, for example, the dynamics of high energy astrophysical plasmas and the quark-gluon plasma created in relativistic heavy-ion collisions.  
Since the seminal work of Israel and Stewart \cite{Israel:1976tn,Israel:1979wp} there have been many papers that have addressed the questions of how to apply and systematically improve relativistic viscous hydrodynamics \cite{Muronga:2001zk,Baier:2006um,Romatschke:2007mq,Baier:2007ix,Dusling:2007gi,Luzum:2008cw,Song:2008hj,El:2009vj,PeraltaRamos:2010je,Denicol:2010tr,Denicol:2010xn,Schenke:2010rr,Schenke:2011tv,Bozek:2011wa,Niemi:2011ix,Niemi:2012ry,Bozek:2012qs,Denicol:2012cn,Jaiswal:2013npa}.  
Recently, a new framework called anisotropic hydrodynamics (aHydro) has emerged for describing the non-equilibrium dynamics of relativistic systems \cite{Florkowski:2010cf,Martinez:2010sc,Ryblewski:2010bs,Martinez:2010sd,Ryblewski:2011aq,Martinez:2012tu,Ryblewski:2012rr,Florkowski:2012as}.

In contrast to conventional viscous hydrodynamical treatments, aHydro does not implicitly rely on an assumption that the system is approximately isotropic in momentum-space.  
Instead, momentum-space anisotropies are built in at leading order by utilizing a spheroidal form  for the one-particle distribution function.  
For systems that are boost invariant and homogeneous in the transverse direction it has been proven that the aHydro framework reduces to second order viscous hydrodynamics in the limit of small anisotropies \cite{Martinez:2010sc}.  
In addition, the framework reproduces the longitudinal free-streaming limit and in all cases the one-particle distribution and the transverse/longitudinal pressures are guaranteed to be positive.  

Given these appealing features, the aHydro framework seems to offer a promising alternative to standard viscous hydrodynamical treatments.  
However, in order to judge the efficacy of the framework it is desirable to have an exactly solvable case with which to compare the various approximations.  
With this in mind, in this Letter we exactly solve the Boltzmann equation for a transversely homogeneous boost-invariant system of massless particles in the relaxation time approximation.
We then compare the resulting exact solutions with aHydro and two different second order viscous hydrodynamics approximations.  

Our starting point is the Boltzmann equation $p^\mu \partial_\mu  f(x,p) = C[f(x,p)]$ where $f(x,p)$ is the one-particle distribution function and $C[f]$ is the collisional kernel.  
We will focus on the case of the relaxation time approximation (RTA) \cite{1954PhRv...94..511B}

\begin{equation}
C[f] = \frac{p_\mu u^\mu}{\tau_{\rm eq}} \biggl[ f_{\rm eq}\Big(p_\mu u^\mu,T(x)\Big) -f(x,p) \biggr],
\label{eq:col-term}
\end{equation}
where $u^\mu$ is the local rest frame four velocity, $\tau_{\rm eq}$ is the relaxation time which may depend on proper time, and $f_{\rm eq}$ is an equilibrium distribution function that may be taken to be a Bose-Einstein, Fermi-Dirac, or Boltzmann distribution.  
The effective temperature $T(\tau)$ appearing in the argument of the equilibrium distribution function is fixed by requiring dynamical energy-momentum conservation \cite{Baym:1984np}.

We focus on boost invariant systems which are homogenous in the transverse direction in which case the dynamical variables only depend on the proper time.  
We additionally specialize to the case of massless particles.
Defining variables $w =  t p_L - z E$ and $v = Et-p_L z$ \cite{Bialas:1984wv,Bialas:1987en,Florkowski:2012ax}, where $p_L$ is the particle momentum along the $z$ direction, the left hand side of the Boltzmann equation can be written simply as $p^\mu \partial_\mu f = (v/\tau) \partial_\tau f$.  
This allows one to solve the RTA Boltzmann equation exactly
\begin{equation}
f(\tau,w,p_\perp) = D(\tau,\tau_0) f_0(w,p_\perp)  
+ \int_{\tau_0}^\tau \frac{d\tau^\prime}{\tau_{\rm eq}(\tau^\prime)} \, D(\tau,\tau^\prime) \, 
f_{\rm eq}(\tau^\prime,w,p_\perp) \, ,  
\label{eq:solf}
\end{equation}
where $\tau_0$ is the initial proper time, $f_0$ is the initial non-equilibrium distribution function, and ${D(\tau_2,\tau_1) = \exp\!\left[-\int_{\tau_1}^{\tau_2} d\tau^{\prime\prime} \, \tau^{-1}_{\rm eq}(\tau^{\prime\prime})\right]}$ is the damping function.
This solution is similar to the one obtained originally by Baym \cite{Baym:1984np}, see also \cite{Heiselberg:1995sh,Wong:1996va}.  
We have generalized it to an arbitrary initial condition at $\tau_0\neq0$ and allowed for the possibility that the equilibration time $\tau_{\rm eq}$ is time dependent.  In the relaxation time approximation one finds $\tau_{\rm eq} = 5 \eta/(T {\cal S})$ where $\eta$ is the shear viscosity, ${\cal S}$ is the entropy density, and $T$ is the effective temperature which we will specify below \cite{Anderson1974466,Czyz:1986mr}.\footnote{We note that when employing the Grad-Israel-Stewart approximation truncated at second order in moments, one finds instead $\tau_{\rm eq} = 6 \eta/(T {\cal S})$.  This is an artifact of the second order truncation.  We will return to this issue later and demonstrate that the value $\tau_{\rm eq} = 6 \eta/(T {\cal S})$ is not in agreement with the exact solution.}
Herein we will assume that $\eta/ {\cal S}$ is time independent. 

Based on Eq.~(\ref{eq:solf}), one can evaluate the energy density via
\begin{equation}
{\cal E}(\tau) = g \int dP \, v^2\,  f(\tau,w,p_\perp)/\tau^2 \, ,
\end{equation}
where $g$ is the degeneracy factor and $dP = 2 \, d^4p \, \delta(p^2) \theta(p^0) = v^{-1} \, dw \, d^2p_T$. Integrating Eq.~(\ref{eq:solf}) one  obtains an integral equation for the energy density
\begin{eqnarray}
\bar{\cal E}(\tau) &=& D(\tau,\tau_0) \,
{\cal R}\big(\xi_{\rm FS}(\tau)\big)\big/{\cal R}\left(\xi_0\right)
\nonumber \\
&& \hspace{1cm}
+ \int_{\tau_0}^{\tau} \! \frac{d\tau^\prime}{\tau_{\rm eq}(\tau^\prime)} \, D(\tau,\tau^\prime) \, 
\bar{\cal E}(\tau^\prime) \, {\cal R}\!\left( \! \left(\frac{\tau}{\tau^\prime}\right)^2 - 1 \right) ,
\label{eq:inteq}
\end{eqnarray}
where ${\bar{\cal E} = {\cal E}/{\cal E}_0}$ is the energy density scaled by the initial energy density, $\xi_0$ is the initial momentum-space anisotropy,
${\xi_{\rm FS}(\tau) = (1+\xi_0)(\tau/\tau_0)^2-1}$, and ${\cal R}(z) = \frac{1}{2} \big[ (1+z)^{-1} + \arctan\!\big(\sqrt{z}\big)/\sqrt{z} \big]$.

\begin{figure}[t!]
\begin{center}
\parbox{10cm}{\includegraphics[width=22cm]{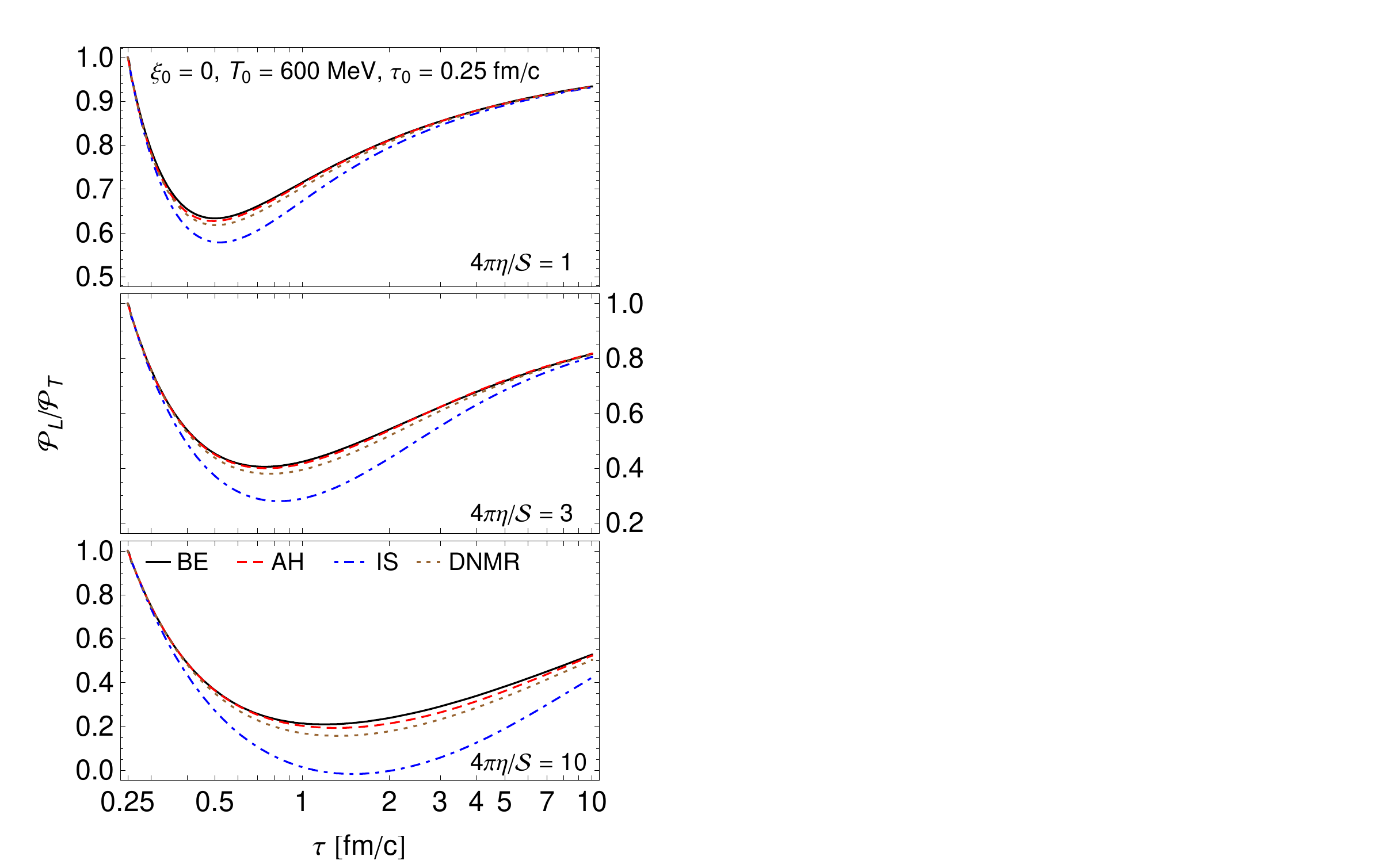}}
\end{center}
\caption{(Color online) Pressure anisotropy as a function of proper time assuming $\xi_0 = 0$ and $T_0 = 600$ MeV at $\tau_0 =$ 0.25 fm/c for $4 \pi \eta/S =$ 1 (top), 3 (middle), and 10 (bottom).  Exact solution (black solid), aHydro (AH) approximation (red long-dashed), Israel-Stewart (IS) approximation (blue dot-dashed), and full second order (DNMR) approximation (brown dotted) \cite{Denicol:2012cn} are compared.}
\label{fig:PLoverPT_600_0}
\end{figure}

\begin{figure}[t!]
\begin{center}
\parbox{10cm}{\includegraphics[width=22cm]{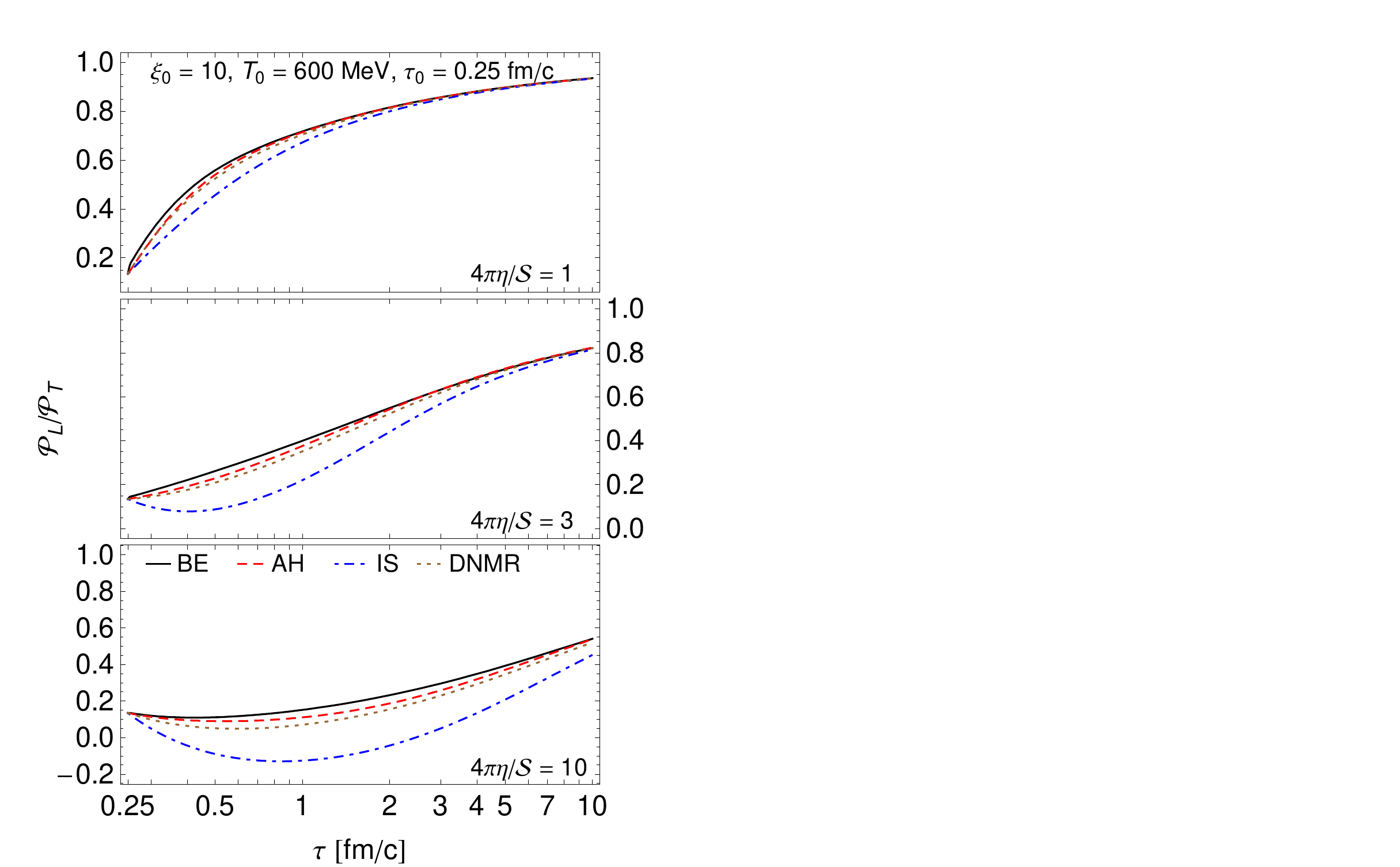}}
\end{center}
\caption{(Color online) Pressure anisotropy as a function of proper time assuming $\xi_0 = 10$ and $T_0 = 600$ MeV at $\tau_0 =$ 0.25 fm/c for $4 \pi \eta/S =$ 1 (top), 3 (middle), and 10 (bottom).  Labeling is the same as in Fig.~\ref{fig:PLoverPT_600_0}.}
\label{fig:PLoverPT_600_10}
\end{figure}

Equation~(\ref{eq:inteq}) can be solved numerically using the iteration method.  
From the resulting energy density, one can solve for the effective temperature via ${\cal E}(\tau) = \gamma \, T^4(\tau)$ where $\gamma$ is a constant which depends on the particular equilibrium distribution function assumed and the number of degrees of freedom.  
The resulting effective temperature allows one to determine the distribution function $f_{\rm eq}$ at all proper times and, with this, the full particle distribution function can be obtained using Eq.~(\ref{eq:solf}).
Additionally, one can determine the number density, longitudinal pressure, and transverse pressure by integrating the distribution function multiplied by $v/\tau$, $w^2/\tau^2$, and $p_T^2/2$, respectively.   

We will compare the resulting numerical solutions with the solution of the aHydro (AH) equations which are obtained by evaluating the zeroth and first moments of Boltzmann equation in the relaxation time approximation assuming a spheroidal form for the distribution function $f_{\rm AH} = f_{\rm eq}\Big(\big((p_\mu u^\mu)^2 + \xi (p_\mu z^\mu)^2\big)/\Lambda^2\Big)$ with $u^\mu=(t,0,0,z)/\tau$ and $z^\mu=(z,0,0,t)/\tau$.  
With this assumption, one obtains~\cite{Martinez:2010sc}
\begin{eqnarray}
\frac{1}{1+\xi} \partial_\tau \xi- \frac{2}{\tau} - \frac{6}{\Lambda} \partial_\tau \Lambda &=& 
\frac{2}{\tau_{\rm eq}^{\rm AH}} \left[ 1 - {\cal R}^{3/4}(\xi) \sqrt{1+\xi} \right] \! ,
\nonumber \\
\frac{{\cal R}'(\xi)}{{\cal R}(\xi)} \partial_\tau \xi + \frac{4}{\Lambda} \partial_\tau \Lambda &=& 
\frac{1}{\tau} \left[ \frac{1}{\xi(1+\xi){\cal R}(\xi)} - \frac{1}{\xi} - 1 \right] \! , \;\;\;\;
\label{eq:ahydro}
\end{eqnarray}
where $\Lambda$ is the transverse temperature and ${\tau_{\rm eq}^{\rm AH} = 5 \eta/(2 \Lambda {\cal S})}$ is the relaxation time.\footnote{The identification of $\tau_{\rm eq}^{\rm AH}$ used herein differs from the prescription specified in Ref.~\cite{Martinez:2010sc} where the scale was set by the effective temperature instead of $\Lambda$.}
We note that in the limit $\xi \rightarrow 0$ one has $\Lambda \rightarrow T$ and ${\tau_{\rm eq}^{\rm AH} = \tau_{\rm eq}/2}$.
The time evolution of $\xi$ and $\Lambda$ is obtained by solving Eqs.~(\ref{eq:ahydro}) and using these one can straightforwardly compute the time dependence of the energy density, transverse pressure, longitudinal pressure, and number density \cite{Martinez:2009ry,Martinez:2010sc}.

\begin{figure}[t!]
\begin{center}
\parbox{10cm}{\includegraphics[width=22cm]{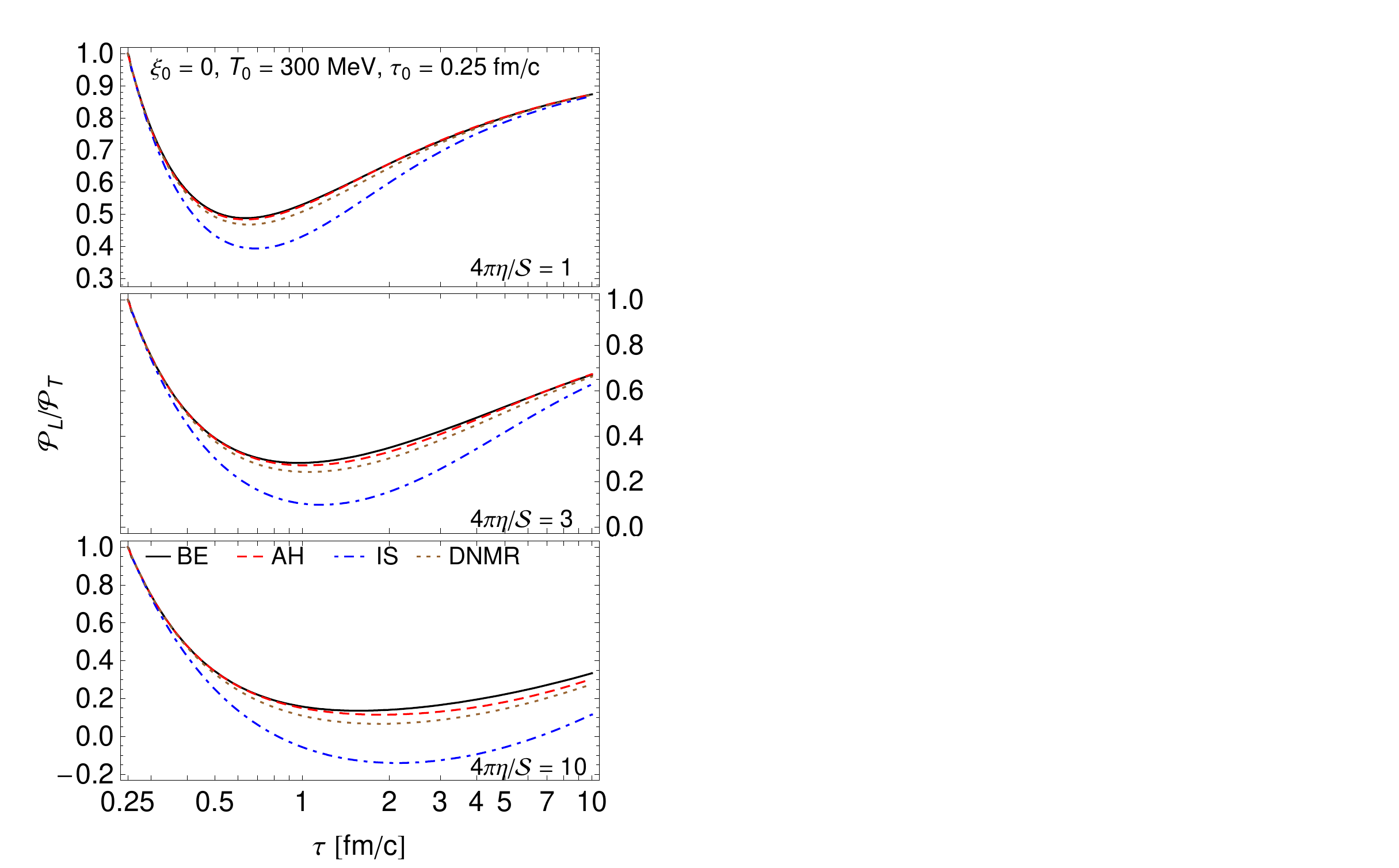}}
\end{center}
\caption{(Color online) Pressure anisotropy as a function of proper time assuming $\xi_0 = 0$ and $T_0 = 300$ MeV at $\tau_0 =$ 0.25 fm/c for $4 \pi \eta/S =$ 1 (top), 3 (middle), and 10 (bottom).  
Labeling is the same as in Fig.~\ref{fig:PLoverPT_600_0}.}
\label{fig:PLoverPT_300_0}
\end{figure}

\begin{figure}[t!]
\begin{center}
\parbox{10cm}{\includegraphics[width=22cm]{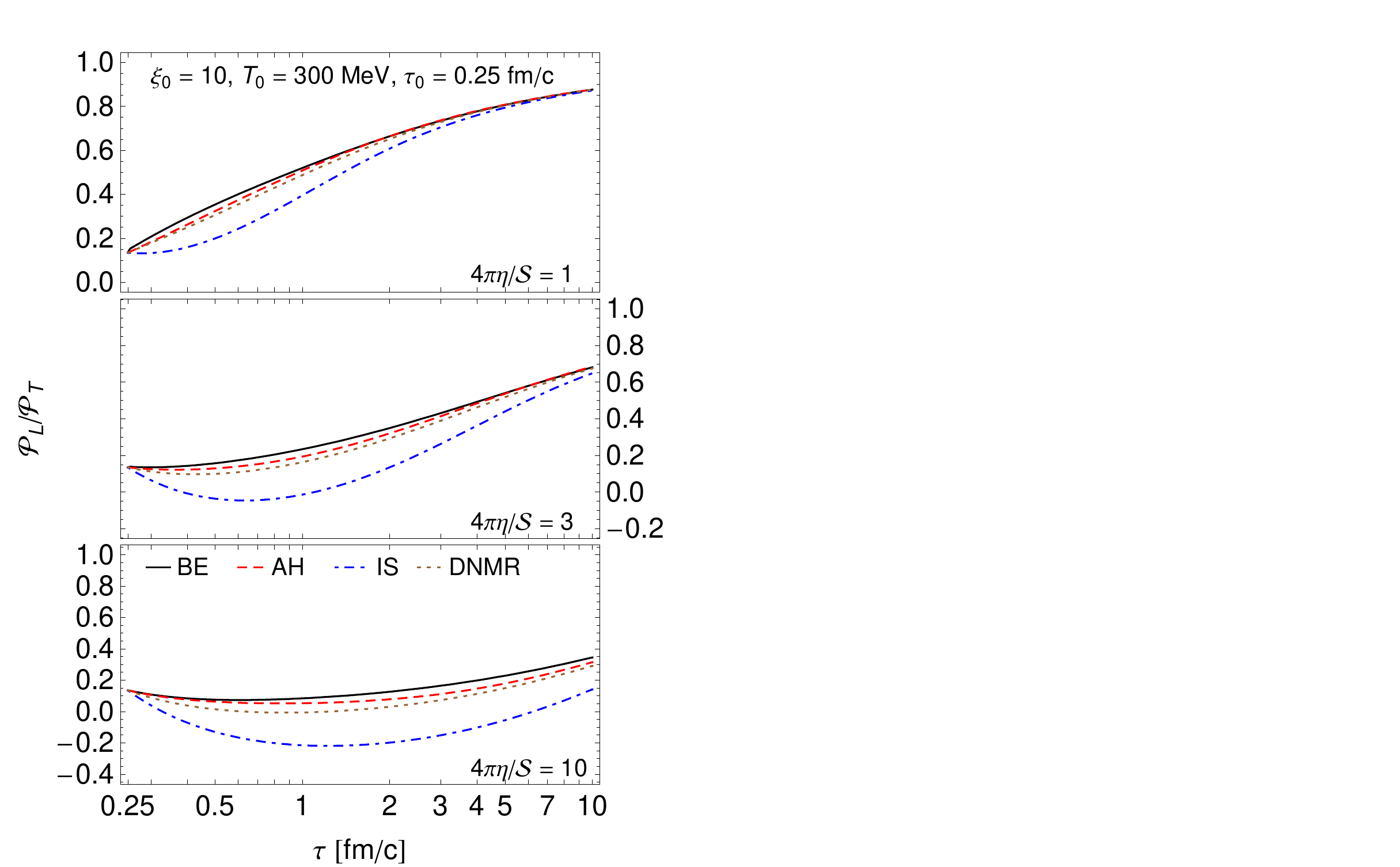}}
\end{center}
\caption{(Color online) Pressure anisotropy as a function of proper time assuming $\xi_0 = 10$ and $T_0 = 300$ MeV at $\tau_0 =$ 0.25 fm/c for $4 \pi \eta/S =$ 1 (top), 3 (middle), and 10 (bottom).  
Labeling is the same as in Fig.~\ref{fig:PLoverPT_600_0}.}
\label{fig:PLoverPT_300_10}
\end{figure}

In addition we will compare the exact solution with two second order viscous hydro prescriptions, both of which can be written compactly as 
\begin{eqnarray}
\partial_\tau {\cal E}&=&-\frac{{\cal E}+{\cal P}}{\tau}+\frac{\Pi}{\tau} \; ,
\nonumber \\
\partial_\tau\Pi&=&-\frac{\Pi}{\tau_\pi}+\frac{4}{3}\frac{\eta}{\tau_\pi\tau}-\beta\frac{\Pi}{\tau}\,,
\label{eq:vhydro}
\end{eqnarray}
where $\Pi$ is the shear and $\tau_\pi = 5 \eta/(T {\cal S})$ is the shear relaxation time. In the majority of the literature, practitioners use $\beta = 4/3$ which we will refer to as the Israel-Stewart (IS) prescription.  
We will also compare the exact solutions with the complete second order treatment from Ref.~\cite{Denicol:2012cn} which, within the relaxation time approximation, gives $\beta = 38/21$.  
We will refer to the second choice as the DNMR prescription.\footnote{Reference~\cite{Jaiswal:2013npa} has also obtained $\lambda = 38/21$ with a different technique.}  In both cases one can compute the transverse pressure via ${\cal P}_T = {\cal P} + \Pi/2$ and the longitudinal pressure via ${\cal P}_L = {\cal P} - \Pi$.  To be consistent with the exact solution and the aHydro approximation we assume an ideal equation of state for the viscous hydrodynamical approximations.

We now turn to our results.  
For all results shown we have assumed that the initial distribution function was spheroidal in form but for the exact solution we do not restrict the distribution function in any way after this point in time.
In Fig.~\ref{fig:PLoverPT_600_0} we show the pressure anisotropy as a function of proper time assuming an initial isotropic plasma with $\xi_0 = 0$ and $T_0 = 600$ MeV at $\tau_0 =$ 0.25 fm/c.  
The three panels show three different assumed values of the shear viscosity to entropy ratio corresponding to $4 \pi \eta/{\cal S} \in \{1,3,10\}$.  
In the figure the aHydro, Israel-Stewart, and full second order viscous hydrodynamics approximations \cite{Denicol:2012cn} are compared with numerical solution of the Boltzmann equation.  
In Fig.~\ref{fig:PLoverPT_600_10} we plot the same quantity for a system possessing an initial momentum-space anisotropy corresponding to $\xi_0 = 10$.  
In Figs.~\ref{fig:PLoverPT_300_0} and \ref{fig:PLoverPT_300_10} we present the pressure anisotropy subject to the same initial conditions and values of $\eta/S$ only changing the initial (effective) temperature to $T_0 = 300$ MeV.

As these figures demonstrate, the aHydro approximation is always closer to the exact solution than the IS and DNMR approximations.  
The IS approximation is the worst approximation to the exact solution in all cases shown and for the case $4\pi\eta/{\cal S} = 10$ it even predicts a negative longitudinal pressure for the majority of the time shown.  
The DNMR approximation represents a significant improvement over the IS approximation; however, we note that if one increases the shear viscosity to entropy ratio even further, the DNMR approximation also predicts negative longitudinal pressures.
Within the aHydro approximation, on the other hand, the pressures are guaranteed to be positive at all times.  
In addition, within aHydro the one-particle distribution function is guaranteed to be positive at all times.

\begin{figure}[t!]
\begin{center}
\includegraphics[width=6.5cm]{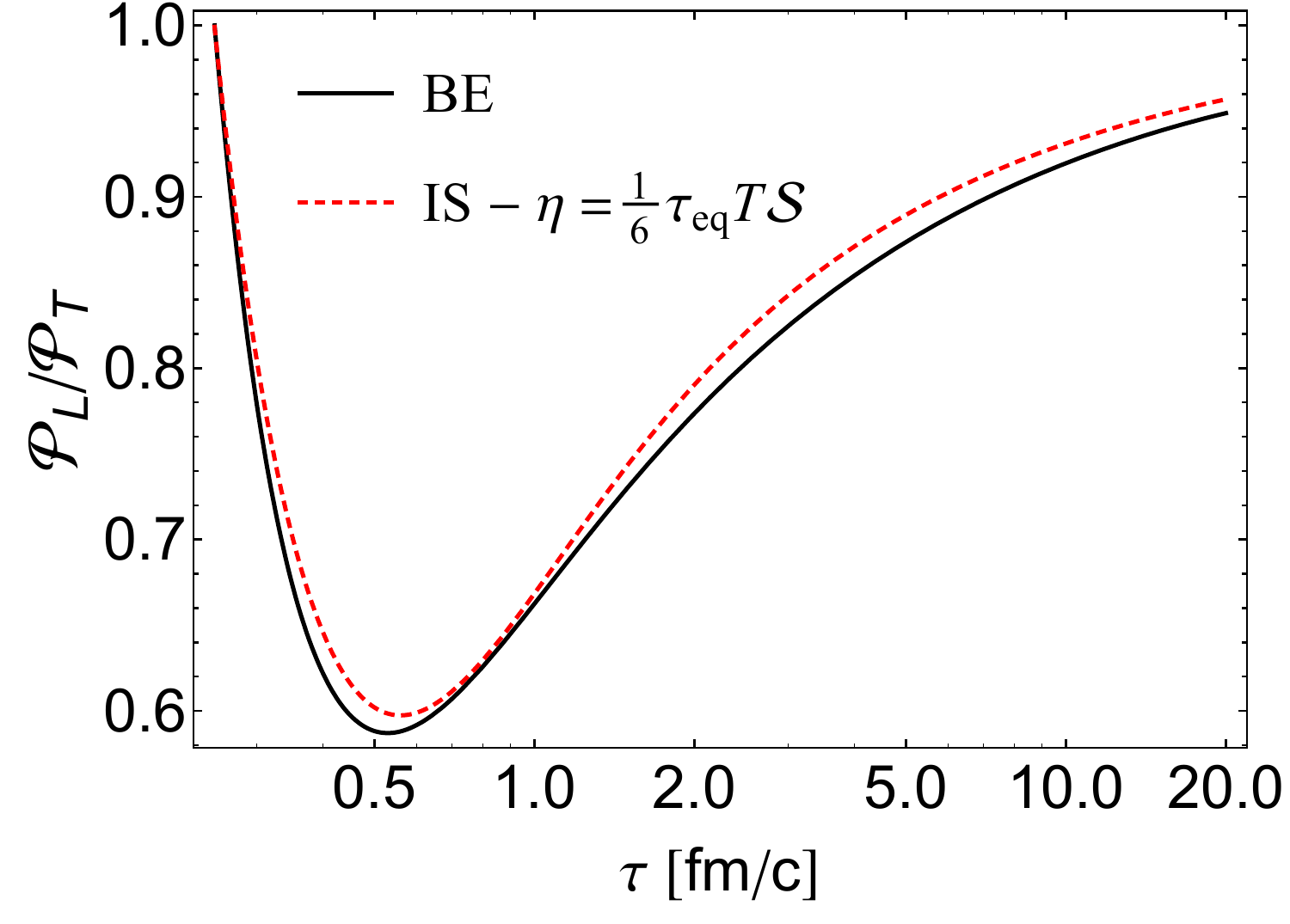}\quad\includegraphics[width=6.5cm]{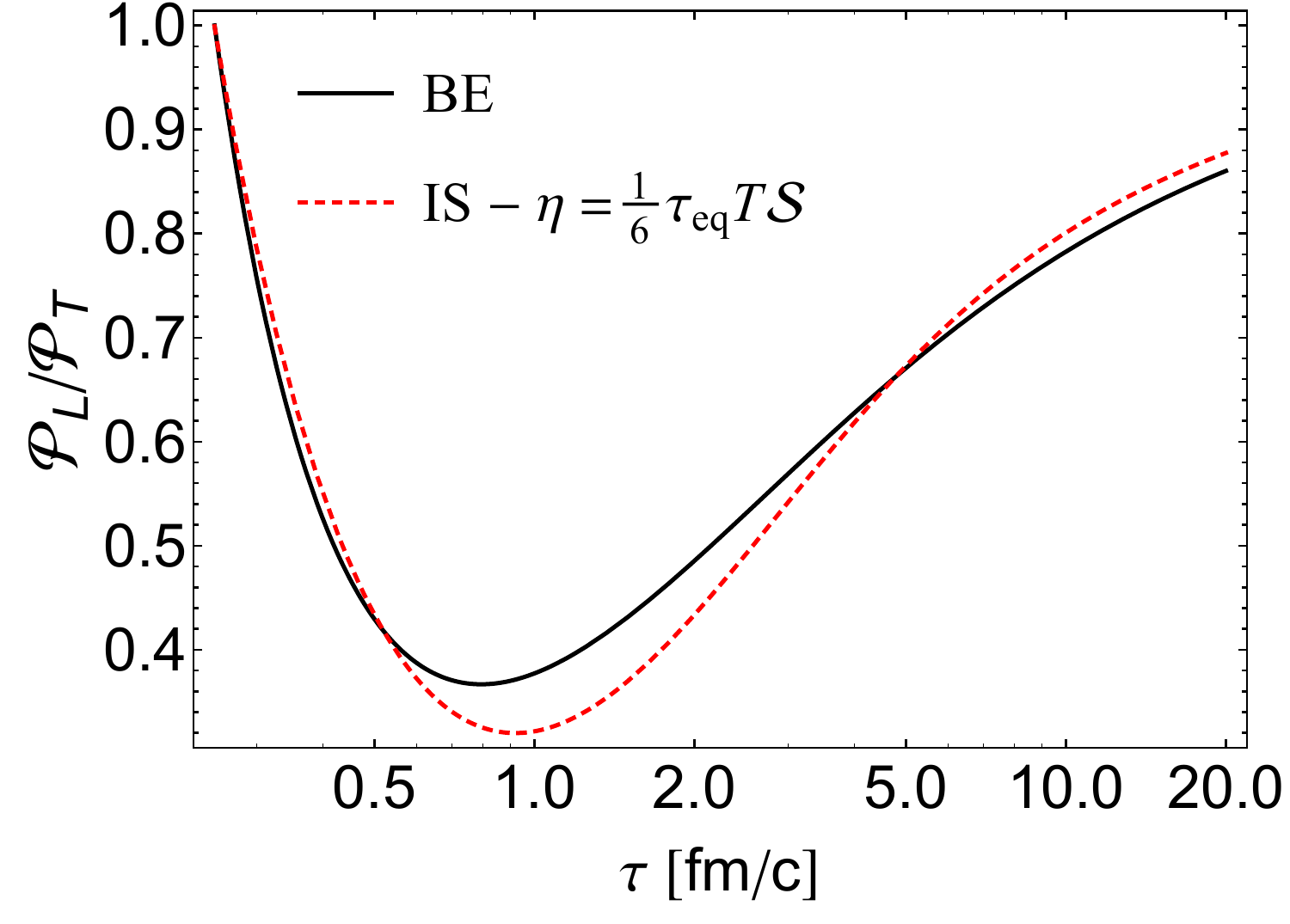}
\end{center}
\vspace{-4mm}
\caption{(Color online) Comparison of the exact solution (BE) with the Israel-Stewart (IS) approximation
assuming $\eta = \tau_{\rm eq} T {\cal S}/6$.
The parameters used were $\xi_0 = 0$ and $T_0 = 600$ MeV at $\tau_0 =$ 0.25 fm/c for (left) $4 \pi \eta/S =$ 1 and (right) $4 \pi \eta/S =$ 3.
}
\label{fig:giscomp}
\end{figure}

As mentioned earlier, if one uses the Grad-Israel-Stewart approximation truncated at second order in moments one erroneously obtains $\tau_{\rm eq} = 6 \eta/(T {\cal S})$ \cite{Romatschke:PC}.  
If one instead uses the Chapman-Enskog method \cite{Anderson1974466,Romatschke:2011qp}, a complete second order Grad expansion \cite{Denicol:2012cn}, or asymptotic expansion without moment expansion \cite{Florkowski:2013lya}, one obtains the correct value of $\tau_{\rm eq} = 5 \eta/(T {\cal S})$.
Whether one obtains $\tau_{\rm eq} = 6 \eta/(T {\cal S})$ or $\tau_{\rm eq} = 5 \eta/(T {\cal S})$ is not specific to 2nd order viscous hydrodynamics as specified in Eqs.~(\ref{eq:vhydro}), but instead is a result of the approximations used when treating the collisional kernel itself.
Since both results are quoted in the literature, in Fig.~\ref{fig:giscomp} we compare the results obtained via exact solution of the relaxation time approximation Boltzmann equation and the Grad-Stewart-Israel approximation assuming $\tau_{\rm eq} = 6 \eta/(T {\cal S})$ with $4 \pi \eta/S =$ 1 (left) and $4 \pi \eta/S =$ 3 (right).
As can be seen from this figure, the Grad-Stewart-Israel approximation with $\tau_{\rm eq} = 6 \eta/(T {\cal S})$ disagrees at early and late times with the exact solution.  
This should be contrasted with Fig.~\ref{fig:PLoverPT_600_0} in which we have used the correct value of $\tau_{\rm eq} = 5 \eta/(T {\cal S})$.  In this case one sees agreement at early and late times with the exact solution.

\begin{figure}[t!]
\begin{center}  w
\parbox{11.5cm}{\includegraphics[width=22cm]{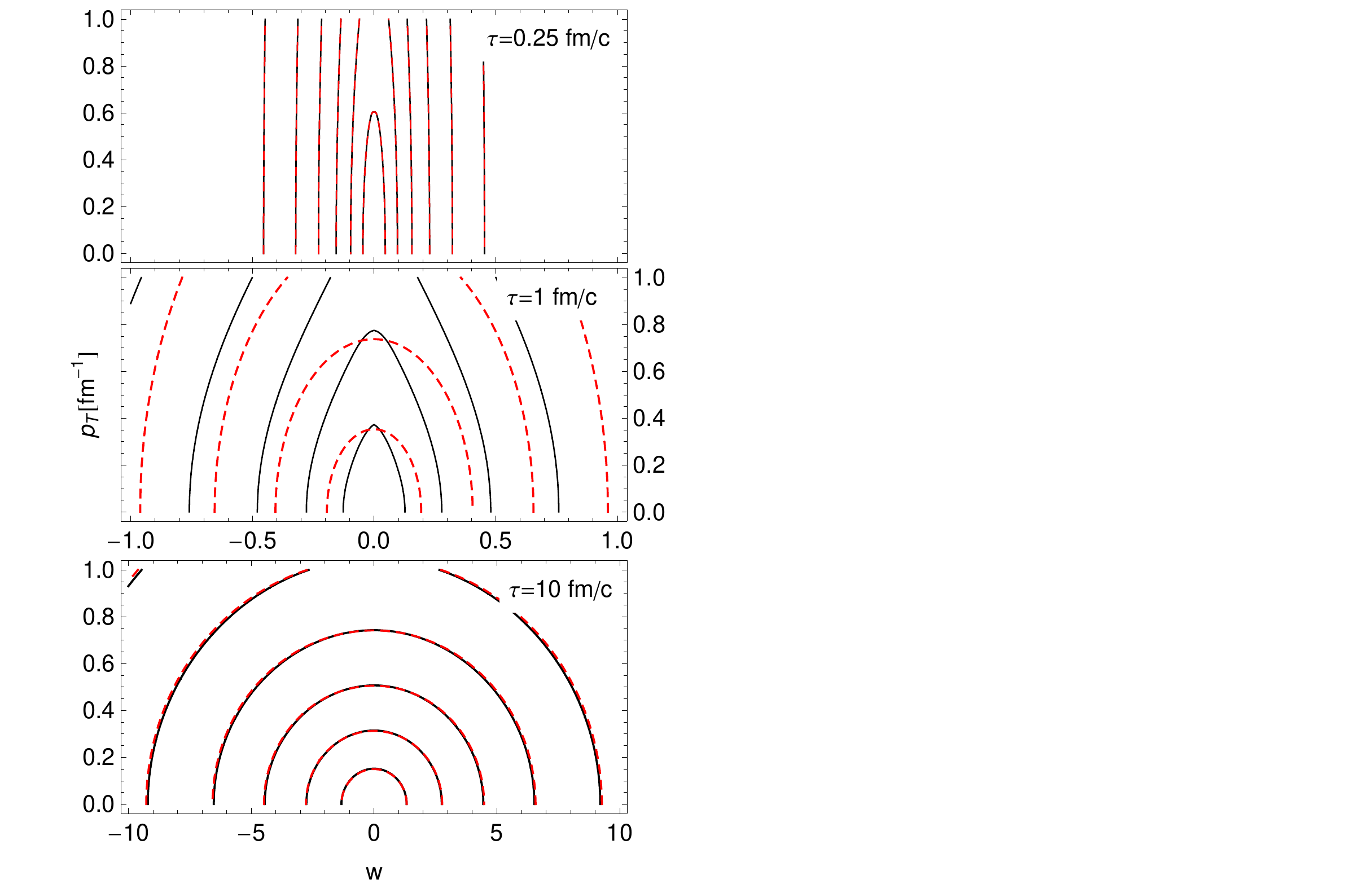}}
\end{center}
\caption{(Color online) Distribution function contour lines resulting from the exact numerical solution (black solid) and the aHydro approximation (red dashed) at three different times $\tau = 0.25$ fm/c (top), 1 fm/c (middle), and 10 fm/c.  
The parameters used where $\xi_0 = 10$ and $T_0 = 600$ MeV at $\tau_0 =$ 0.25 fm/c for $4 \pi \eta/S =$ 3.
}
\label{fig:G_600_10_3over4pi_t1}
\end{figure}

\begin{figure}[t!]
\begin{center}
\includegraphics[width=10.5cm]{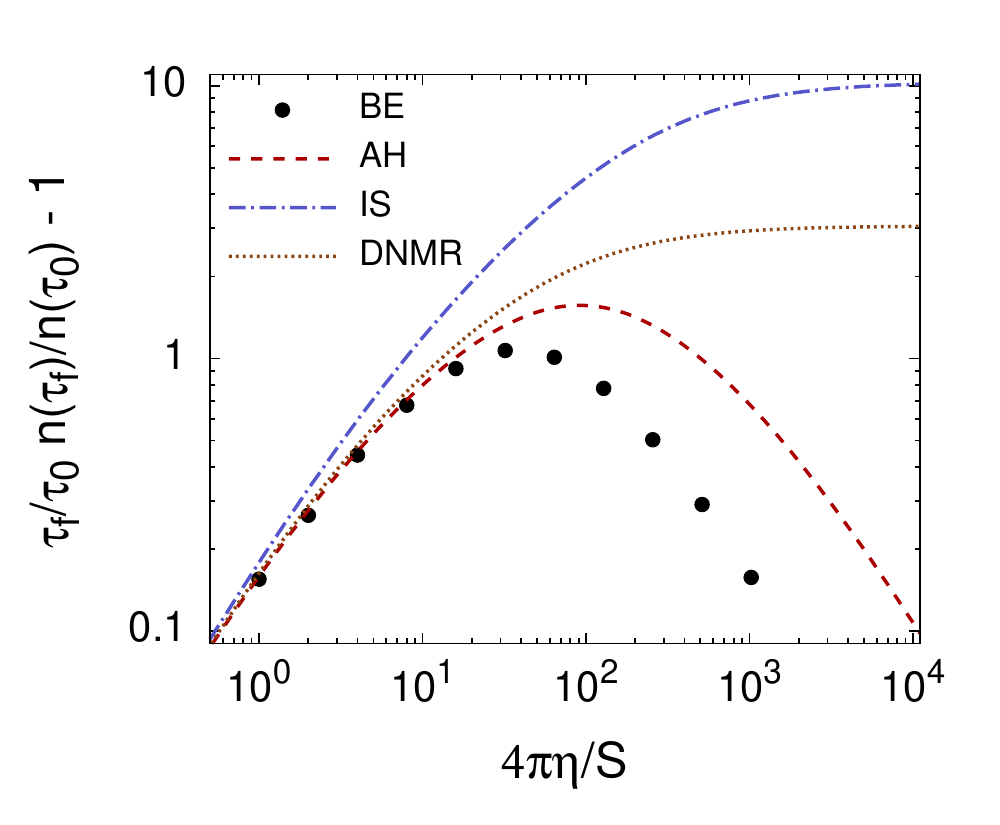}
\end{center}
\vspace{-4mm}
\caption{(Color online) Particle production measure $\Delta_n = \tau_f/\tau_0 \,n(\tau_f)/n(\tau_0) -1$ as a function of $4 \pi \eta/{\cal S}$.
Exact solution (black points), aHydro (AH) approximation (red long-dashed), Israel-Stewart (IS) approximation (blue dot-dashed), and full second order (DNMR) approximation (brown dotted) \cite{Denicol:2012cn} are compared.
}
\label{fig:entropy-gen}
\end{figure}

In Fig.~\ref{fig:G_600_10_3over4pi_t1} we show three contour plots which compare the evolution of the one-particle distribution function obtained from the numerical solution of the Boltzmann equation with the aHydro approximation.
As this figure shows, the exact solution shows deviations from the spheroidal form of aHydro at early times.  
The magnitude of these deviations decreases as the shear viscosity to entropy ratio decreases.  
However, as Figs.~\ref{fig:PLoverPT_600_0}-\ref{fig:PLoverPT_300_10} demonstrate, moments of the distribution function are described quite well.  
We note that the shape of the contours shown in Fig.~\ref{fig:G_600_10_3over4pi_t1} for the exact result suggest that it may be more effective to describe the distribution function in terms of a linear superposition of two spheroidal forms, one which is related to the free streaming contribution coming from the initial distribution function (first term in Eq.~(\ref{eq:solf})) and the other which describes the late-time dynamics.

In Fig.~\ref{fig:entropy-gen} we show the particle number generation measured via $\Delta_n \equiv \tau_f/\tau_0 \,n(\tau_f)/n(\tau_0) -1$ where $\tau_f$ is the freeze-out time defined by when the effective temperature drops below $T_f = $ 150 MeV starting with isotropic $\xi_0 = 0$ initial conditions and a temperature of 600 MeV at $\tau_0 = $ 0.25 fm/c.
For the IS and DNMR approximations, the number density is defined via $n \propto T^3$ where $T$ is the effective temperature obtained via the fourth root of the energy density.  
In the case of the RTA and aHydro results, the number density is computed from the underlying distribution function.
In Fig.~\ref{fig:entropy-gen}, we compare $\Delta_n$ obtained via numerical solution of the Boltzmann equation, the aHydro approximation, the IS viscous hydrodynamical equations, and the DNMR viscous hydrodynamical equations.
On physical grounds one expects this ratio to vanish in the ideal hydrodynamical limit ($\eta/S \rightarrow 0$) and the free-streaming limit ($\eta/S \rightarrow \infty$).
As this figure demonstrates both the IS and DNMR approximations predict that $\Delta_n$ is a monotonically increasing function of the shear viscosity to entropy ratio.  
On the other hand, although the aHydro framework has too much particle production at large $\eta/S$, it has the right qualitative behavior.

In conclusion, we have presented an exactly solvable case in which the Boltzmann equation can be straightforwardly solved numerically.  
The resulting \mbox{0+1-dimensional} RTA solution, while not comprehensive, constitutes a toy model that can be used to test the accuracy of different dynamical approximation schemes for a variety of initial conditions, values of the shear viscosity, etc.  
In the case considered herein, we found that the aHydro approximation is closer to the exact RTA solution than both the IS and DNMR approximations in all cases.
This is remarkable since the aHydro equations themselves are essentially zeroth order in a general anisotropic expansion of the one-particle distribution function, whereas the other approximations are at their respective second order of approximation.  
If deviations from the exact spheroidal form are taken into account, one expects the quality of the aHydro approximation to further improve.  
We note in closing that the exact solutions obtained herein can perhaps be used to create a better approximation framework which takes into account such deviations in a systematic manner.  


\section*{Acknowledgments}
We thank G.~Denicol and P. Romatschke for discussions.  
W.F. was supported in part by the Polish National Science Center with decision no. DEC-2012/06/A/ST2/00390.
R.R. was supported by a Polish National Science Center grant with decision No. DEC-2012/07/D/ST2/02125 and the Foundation for Polish Science.

\bibliographystyle{utphys}
\bibliography{rta}

\end{document}